\documentclass[pra,twocolumn,showpacs,superscriptaddress,floatfix]{revtex4-1}
\usepackage[caption=false]{subfig}
\usepackage{graphicx}
\graphicspath{ {} }
\usepackage{amsmath,graphicx}
\usepackage{epsfig}
\usepackage{amsmath,amssymb,mathrsfs,esint}
\usepackage{gensymb}
\usepackage[bookmarks=true,
   colorlinks=true,
   linkcolor=blue,
   urlcolor=blue,
   citecolor=blue,
   bookmarks=true,
   hyperindex=true
]{hyperref}
\usepackage{natbib}
\usepackage{relsize}
\usepackage{float}
\usepackage{dsfont}

\usepackage{soul}

\newcommand{\be}{\begin{equation}}
\newcommand{\ee}{\end{equation}}
\newcommand{\norm}[1]{\left\lVert#1\right\rVert}

\begin{document}

\title{Quasi-conserved quantities in the perturbed XXX spin chain } 

\author{Savvas Malikis}
\email[]{malikis@lorentz.leidenuniv.nl}
\affiliation{Instituut-Lorentz, Universiteit Leiden, Leiden, The Netherlands}

\author{Denis V. Kurlov}
\affiliation{Russian Quantum Center, Skolkovo, Moscow 143025, Russia}

\author{Vladimir Gritsev}
\affiliation{Institute for Theoretical Physics Amsterdam, Universiteit van Amsterdam, Amsterdam, The Netherlands}
\affiliation{Russian Quantum Center, Skolkovo, Moscow 143025, Russia}

\begin{abstract}
We consider the isotropic spin-$1/2$ Heisenberg spin chain weakly perturbed by a local translationally- and $SU(2)$-invariant perturbation. Starting from the local integrals of motion of the unperturbed model, we modify them in order to obtain quasi-conserved integrals of motion (charges) for the perturbed model. Such quasi-conserved quantities are believed to be responsible for the existence of the prethermalization phase at intermediate timescales. We find that for a sufficiently local perturbation only the first few integrals of motion can be promoted to the quasi-conserved charges, whereas higher-order integrals of motion do not survive. 
\end{abstract}

\maketitle 

\section{Introduction}

In classical mechanics there is a well understood distinction between integrable and non-integrable systems, as well as between their long-time dynamics. Namely, a generic non-integrable system typically exhibits an ergodic behavior, leading to a chaos,
 whereas integrable systems are non-ergodic and their phase space trajectories are confined to some subregions of the phase space due to the existence of many conserved quantities. 
Moreover, there is a result of tremendous importance, the Kolmogorov-Arnold-Moser (KAM) theorem, which ensures that classical integrable systems under a weak integrability-breaking perturbation are stable for a sufficiently long time \cite{arnold1,arnold2,kolmogorov1,moser1,moser2}. 

Extending the KAM theorem to the quantum case is a long-standing problem. Although recent findings demonstrate some progress in this direction \cite{quantumkam}, a complete understanding is missing and there are numerous open questions.
In part this is due to the fact that in the quantum case even the very definition of integrability is subtle~\cite{quantumint}.

A widely accepted criterion for quantum integrability is that if a Hamiltonian $H_0$ is integrable, then there exists a large number of extensive, functionally-independent, and mutually commuting conserved quantities (charges):
\be \label{local_charges}
	\left[ H_0, Q_j \right] = \left[ Q_k, Q_j \right] = 0.
\ee
Importantly, the conserved charges $Q_j$ are assumed to be local, in a sense that they are given by the sums of operators with a finite support.

Just like in the classical case, the long time dynamics is very different for integrable and non-integrable systems (to be precise, we do not consider here systems that exhibit Anderson or many-body localization).
Non-integrable systems thermalize according to the Eigenstate Thermalization Hypothesys (ETH), which means that the total isolated system acts a heath bath for its own subsystems. This leads to the spread of entanglement over the whole system, such that in the long time limit it is impossible to retrieve any information about the initial state using only local measurements \cite{srendicki, deutschfirst}.
On the contrary, in integrable systems, thermalization is very different. It is described by the generalized Gibbs ensemble (GGE), which takes into account that there are many other conserved quantities apart from the total energy and the number of particles, as it is for the standard grand-canonical ensemble \cite{EF-rev,EF1,poz1}. Moreover, it has been shown that to accurately describe thermalization of integrable systems one should extend the GGE by including not only the local conserved charges, as in Eq. (\ref{local_charges}), but also the {\it quasi-local} ones \cite{quasilocal1,quasilocal2,quasilocalgge,quasilocalgge2}.

Then, in the spirit of the KAM theorem, one may ask what will happen if a quantum integrable system is slightly perturbed away from integrability. What kind of thermalization will it exhibit?
Naively, one would expect that nearly-integrable quantum systems simply thermalize following the ETH.
However, it is widely believed that such systems also exhibit a different, the so-called prethermal, behavior at intermediate times~\cite{prethermalization, prethernearin1, prethermnearin2}. Different studies \cite{pretherml1,pretherml2,pretherml3} suggest that the eventual thermalization occurs at much later times~$t_{\text{th}}\sim \lambda^{-2}$, where~$\lambda \ll 1$ is the strength of the perturbation, and the scaling can be understood from the Golden Rule. Moreover, it is believed that this prethermal phase should be described by some effective GGE \cite{prethgge}. 
Therefore, it is natural to ask what are the charges that define this effective GGE in the prethermal phase. Clearly, since this effective description is only valid at times~$t \lesssim \lambda^{-2}$, these charges can be only {\it quasi}-conserved with the accuracy~${\cal O}(\lambda^2)$. 
In other words, since the exact conservation laws of the unperturbed system constrain the dynamics of an integrable system, one can expect that the dynamics of a perturbed system 
should be restricted by the quasi-conserved charges. 

This idea is also supported by the developments in the context of the slowest operators~\cite{slow1,slow2}. Indeed, for an operator~$O$ that commutes with a Hamiltonian~$H$, the time evolution~$e^{iHt}Oe^{-iHt}$ is trivial. In terms of the quantum information language this means that the information encoded in $O(0)$ does not spread. On the contrary, if~$[O(0),H]\neq 0$, the typical timescale of information spreading is inversely proportional to the norm of~$[O(0), H]$, as follows from the Baker-Campbell-Hausdorff formula. Thus, to slow down the spread of the quantum information one needs to suppress (at least) the first order term in the Baker-Campbell-Hausdorff expansion.

Let us mention that the search for quasi-conserved charges can be linked with an old problem in functional analysis, related to almost commuting matrices \cite{math1} and
explicitly stated by Halmos in \cite{Halmos}.
This long-standing question ``when two almost commuting matrices are close to matrices that exactly commute" was answered by H.~Lin~\cite{Lin}.
More precisely, Lin showed that given~$\epsilon>0$ there exists $\delta>0$ such that if $N\times N$ matrices $A,B$ are Hermitian, with $\parallel AB-BA\parallel<\delta$ and $\parallel A\parallel,\parallel B\parallel\leq1$, then there exists commuting Hermitian $N\times N$ matrices~$X,Y$ such that~$\parallel A-X\parallel+\parallel B-Y\parallel<\epsilon$. Here,~$\parallel \cdot \parallel$ is the usual operator norm. Importantly, $\delta=\delta(\epsilon)$ {\it does not} depend on the dimension $N$. Recently, Hastings obtained an explicit estimate $\epsilon(\delta)\sim \delta^{1/5}$, where the exponent may depend on the choice of the operator norm~\cite{Hastings}. Quite remarkably, the question whether one can find triples of almost commuting matrices has generically a negative answer \cite{triple}. This perhaps is connected to what we found in this paper: it is probably impossible to construct a set of higher-order quasi-commuting quasi-conserved charges.    
A similar story about {\it unitary} matrices is more involved~\cite{math-uni1, note}. There, the existence of almost commuting unitary matrices have some topological obstructions given by the so-called Bott indices. There is an extensive mathematical literature on the subject, see, e.g., Ref.~\cite{math-uni2}.   
This may suggest that an analogue of the KAM theory for quantum systems can not be generically defined.
Finally, we would like to mention that there is a somewhat related research direction in the context of AdS/CFT correspondence, which deals with the so-called long-range deformed spin chains~\cite{longrange1, longrange2} and $T \bar T$-deformations~\cite{Marchetto2020, Pozsgay2020}. However, these studies deal with the deformations that preserve integrability to all orders in the perturbation strength and thus differ from the present work.

This paper is devoted to the search for quasi-conserved quantities in a quantum spin chain weakly perturbed away from integrability. The rest of the paper is organized as follows. In Section~\ref{S:model_boost} we describe the model and discuss some general properties of the exact conserved charges that are present in the absence of the perturbation. In Section~\ref{S:quasi_charges} we present an ansatz for the quasi-conserved charges and introduce a measure for their quality. Finally, in Section~\ref{S:results} we demonstrate our numerical findings, conclude, and formulate some open questions for future research.

\section{The model and the boost operator} 
\label{S:model_boost}

We now proceed with the description of the specific model that we are going to deal with.
Consider a Hamiltonian 
\be \label{H_tot}
	H_{\lambda} = H_0 + \lambda H_1,
\ee
where $H_0$ is an integrable part, $H_1$ is an integrability-breaking perturbation, and $\lambda > 0$ is a numerical parameter characterizing the perturbation strength. We assume~$\lambda \ll 1$, such that the perturbation is weak.
For the unperturbed system, we take a spin-$1/2$ isotropic Heisenberg spin chain (XXX model) on a one-dimensional lattice of $N$ sites:
\be \label{H_0}
	H_0 = \sum_{j =1}^N h^{(0)}_j =  \sum_{j} J\, \vec{\sigma}_j\cdot\vec{\sigma}_{j+1},
\ee
where $\vec{\sigma}_{j}$ is the vector of Pauli matrices and $J$ is an exchange constant. It is well known that~$H_0$ is integrable, the exact spectrum and the eigenstates can be found using the Bethe ansatz \cite{bethe}, and one has a large number of conserved charges.
Let us now break the integrability by a perturbation of the following form:
\be \label{H_1}
	H_1 = \sum_{j =1}^N h^{(1)}_j =  \sum_{j} J\, \vec{\sigma}_j\cdot\vec{\sigma}_{j+2},
\ee
which is nothing other than the next-to-nearest neighbor Heisenberg interaction.
In what follows we put $J = 1$ and impose periodic boundary conditions. 
Let us mention that both the unperturbed Hamiltonian~$H_0$ and the perturbation~$H_1$ are translationally- and $SU(2)$-invariant. Also, both are the sums of local operators with the support on two and three sites (for $H_0$ and $H_1$, respectively).
We note in passing that in the limit $\lambda \to \infty$ the Hamiltonian~(\ref{H_tot}) is equivalent to a pair of decoupled XXX models (on even and odd sites), and integrability is restored. However, we do not consider the case of large $\lambda$ and restrict ourselves to $\lambda \ll 1$.
Before we turn to the problem of constructing the quasi-conserved quantities for the perturbed Hamiltonian~(\ref{H_tot}), let us briefly summarize the most important properties of the charges conserved by the integrable Hamiltonian~$H_0$.

 Local conserved charges [as those given in Eq.~(\ref{local_charges})] can be generated iteratively starting from $Q_2 \equiv H_0$ [by convention, $Q_1$ is the total magnetization] and using the following relation~\cite{grabowski1, grabowski2, grabowski3}:
\be \label{Q_sequence}
	Q_{n+1} = i [B,Q_n],
\ee
where $B$ is the so-called boost operator, which reads
\be \label{B_XXX}
    B = \sum_{j} j \, h^{(0)}_j = \sum_{j} j \, \vec{\sigma}_j\cdot\vec{\sigma}_{j+1}.
\ee
Thus, the boost operator acts  as a ladder operator in the space of conserved charges of the model. 
Importantly, every next charge has a larger support as compared to the previous one. For the XXX model, the $n$-th charge $Q_n$ is a sum of operators with a support on $n$ sites. The explicit form of the first few charges can be found in Ref.~\cite{grabowski2}. 
For illustrative purposes, here we present the expression for $Q_3^{(0)}$:
\be \label{Q3}
	Q_3 = \sum_j \vec{\sigma}_{j} \cdot \left( \vec{\sigma}_{j+1} \times \vec{\sigma}_{j+2} \right),
\ee
which has a support on $n=3$ sites.
We emphasize once again that $B$ can only generate the {\it local} charges, whereas the Hamiltonian~(\ref{H_0}) also possesses the quasi-local ones~\cite{quasilocal1}.  To our knowledge, a corresponding boost operator that can generate quasi-local charges has not been found. 

Let us now turn on the perturbation (\ref{H_1}), such that the total Hamiltonian is $H_{\lambda}$ as given by Eq.~(\ref{H_tot}), and~$\lambda \ll 1$. The quantities~$Q_n$ are no longer conserved, since they do not commute with $H_{\lambda}$. Neither they are quasi-conserved, since $\norm{ [H_{\lambda}, Q_n] } \propto\lambda$. Hence, they change significantly over times much shorter than $t_{\text{th}} \sim \lambda^{-2}$ and can not govern the dynamics in the pre-thermal phase.


\section{Quasi-conserved charges}
\label{S:quasi_charges}

We now proceed with looking for the quasi-conserved quantities that survive during the pre-thermal phase up to times $\sim \lambda^{-2}$. This simply means that we are looking for a set of operators $\tilde{Q}_n$ that satisfy
\be \label{comm_norm_scaling}
	\norm{ [ H_{\lambda}, \tilde{Q}_n ] } \propto \lambda^b, \qquad b \geq 2,
\ee
and commute with each other with the accuracy ${\cal O}(\lambda^2)$.
Further, since the Hamiltonian $H_{\lambda}$ has the translational and $SU(2)$ symmetries, we require that the quasi-conserved charges $\tilde{Q}_n$ possess these symmetries as well. 
In analogy with the integrable case, we identify the second charge with the Hamiltonian, i.e., $\tilde{Q}_2 = H_{\lambda}$ [note that $\tilde Q_1 = Q_1$, as the total magnetization is conserved by $H_{\lambda}$]. This gives us the relation
\be
	\tilde{Q}_2 = Q_2 + \lambda H_1.
\ee
Since the perturbation is weak, it is natural to expect that similar relations should hold for higher charges as well.
Therefore, we make an ansatz
\be \label{quasiQ_ansatz}
	\tilde{Q}_{n}^{({\cal M})} = Q_n + \sum_{m = n}^{ {\cal M} } \delta Q_{n}^{(m)}, 
\ee
where $\delta Q_n^{(m)}$ is an operator that is a sum of terms having the support on $m$ sites. We introduce a cutoff ${\cal M}$ on the maximal support, because we are interested in local quasi-conserved charges. However, we have ${\cal M} > n$, i.e., the maximal support of~$\tilde Q_{n}^{({\cal M})}$ is larger than that of~$Q_n$. This is a reasonable assumption because the perturbed Hamiltonian $H_{\lambda}$ itself has a greater support than~$H_{0}$. Moreover, as we already mentioned in the Introduction, even for the integrable $H_0$ the quasi-local charges (that have an infinite support) play an important role.

Let us now discuss the structure of $\delta Q_{n}^{(m)}$ from Eq.~(\ref{quasiQ_ansatz}) in more detail. First of all, translational invariance allows us to express it in the following form:
\be \label{deltaQ}
	\delta Q_{n}^{(m)} = \sum_j \sum_{ \{ \ell_s \}} c_{n}^{(m)}\left( \{ \ell_s \} \right) {\cal Q}_{j}^{(m)} \left( \{ \ell_s \} \right),
\ee
where $c_{n}^{(m)}\left( \{ \ell_s \} \right)$ are numerical coefficients, and an operator ${\cal Q}_{j}^{(m)}\left( \{ \ell_s \} \right)$ has a support on $m$ sites from $j$ to $j+m-1$. It acts non-trivially on $k$ sites ($2 \leq k \leq m$), specified by the set $\{\ell_s\}$:
\be \label{sites}
	j + \ell_0, \; j + \ell_1, \; j + \ell_2 , \ldots ,\; j + \ell_{k-1},
\ee
\\
where $\ell_0 \equiv 0$ and $\ell_{k-1} \equiv m -1$ are fixed, and the rest of positive integers $\ell_s$ are sorted and take the values $1 \leq \ell_1 < \ell_2 < \ldots < \ell_{k-2} \leq m - 2$. Summation over $\{\ell_s\}$ in Eq.~(\ref{deltaQ}) takes into account that there are $\sum_{k=2}^{m} \binom{m-2}{k-2} = 2^{m-2}$ possible operators of such form. However, from these terms we do not include into $\delta Q_n^{(n)}$ those ones that are already present in $Q_n$, i.e., we put to zero the corresponding coefficients $c_n^{(n)}(\{l_s\})$. For instance, for $\delta Q_3^{(3)}$ we have $c_3^{(3)}(\{0,1,2\}) = 0$.

Second, due to the $SU(2)$ symmetry, ${\cal Q}_j^{(m)}(\{ \ell_s \})$ in Eq.~(\ref{deltaQ}) can be expressed in the basis of $SU(2)$-invariant tensor products of $k$ spin-$1/2$ operators. For instance, for $k=2$ this is
$\sum_{\alpha_1, \alpha_2} \delta_{\alpha_1, \alpha_2} \sigma_{j_1}^{\alpha_1} \, \sigma_{j_2}^{\alpha_2}$,
where $\delta_{\alpha_1, \alpha_2}$ is the Kronecker symbol, $\sigma^{\alpha_i}_{j_i}$ are the Pauli matrices acting on the site $j_i$, and $\alpha_i = 1,2,3$.
We discuss the basis of $SU(2)$-invariant operators in more detail in Appendix~\ref{SU2_inv_basis}. For the present purposes it is only important that such basis can be constructed, and all quasi-conserved quantities are thus given by Eqs. (\ref{quasiQ_ansatz}) and~(\ref{deltaQ}). 

Then, treating the coefficients $c_{n}^{(m)}\left( \{ \ell_s \} \right)$ as variational parameters, we may tune them in such a way that the commutator $[H_{\lambda}, \tilde Q_n^{({\cal M})}]$ has the smallest possible norm. Thus, we introduce the following dimensionless quantity:
\be \label{non_commutativity_measure}
	    \mathcal{L}[H_{\lambda}, Q] \equiv 2^{N/2} \sqrt{N} \min_{ \{ c_{n}^{(m)}\left( \{ \ell_s \} \right) \} }\frac{ \norm{ [H_{\lambda},Q] } }{ \norm{H_{\lambda }} \norm{ Q} },
\ee
where $\norm{X} \equiv \sqrt{\text{Tr}(X^{\dag}X)}$ is the Frobenius norm, which we choose for computational convenience, and the factor of $2^{N/2} N$ is included to make $\mathcal{L}[H_{\lambda}, Q]$ independent of the system size. Minimization is performed over all coefficients $c_{n}^{(m)}\left( \{ \ell_s \} \right)$ entering Eq.~(\ref{deltaQ}). Note that by taking the $SU(2)$-symmetry into account, we reduce the number of variational parameters from $3 \times 4^{{\cal M} - 1}$ to $\sum_{m=2}^{{\cal M}} 2^{m-2} = 2^{{\cal M}-1} -1$, achieving a square root improvement.

\begin{widetext}

\begin{figure}[H]
\subfloat{
  \includegraphics[width=0.49\linewidth]{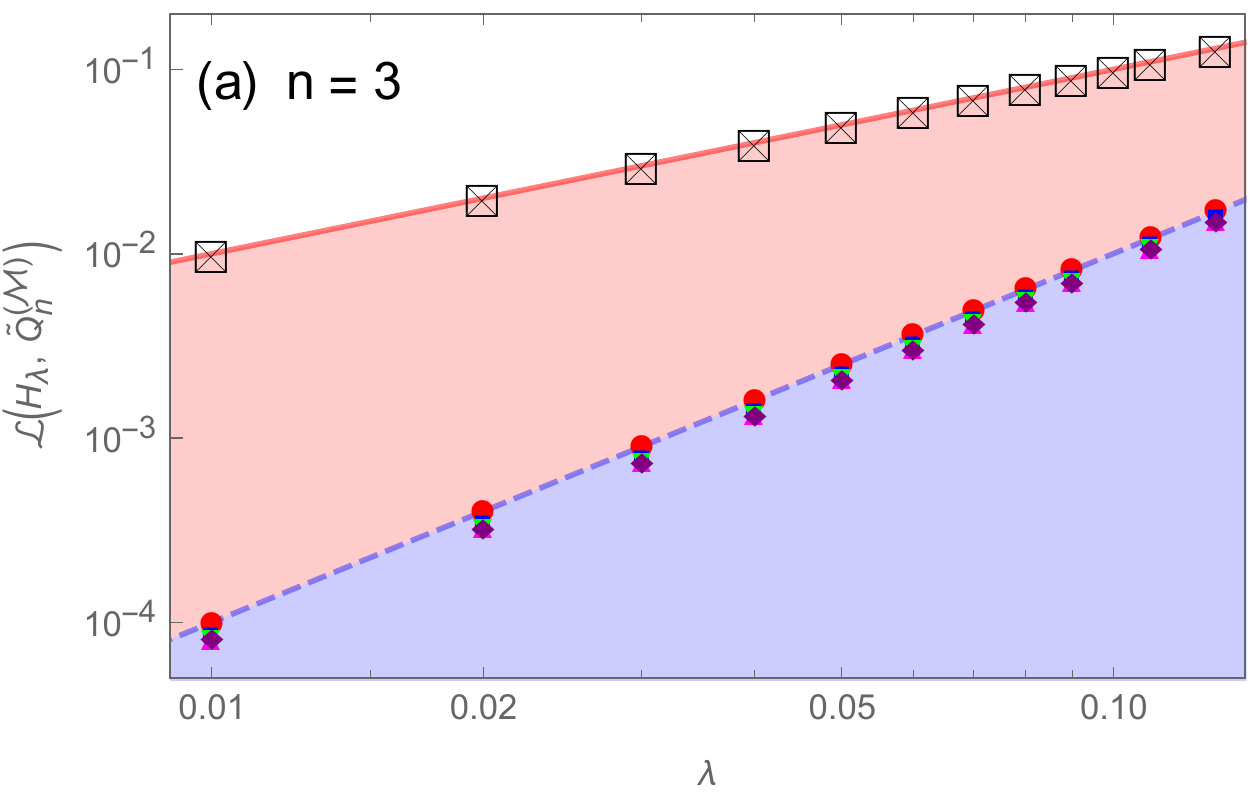}
}\hfill
\subfloat{
  \includegraphics[width=0.49\linewidth]{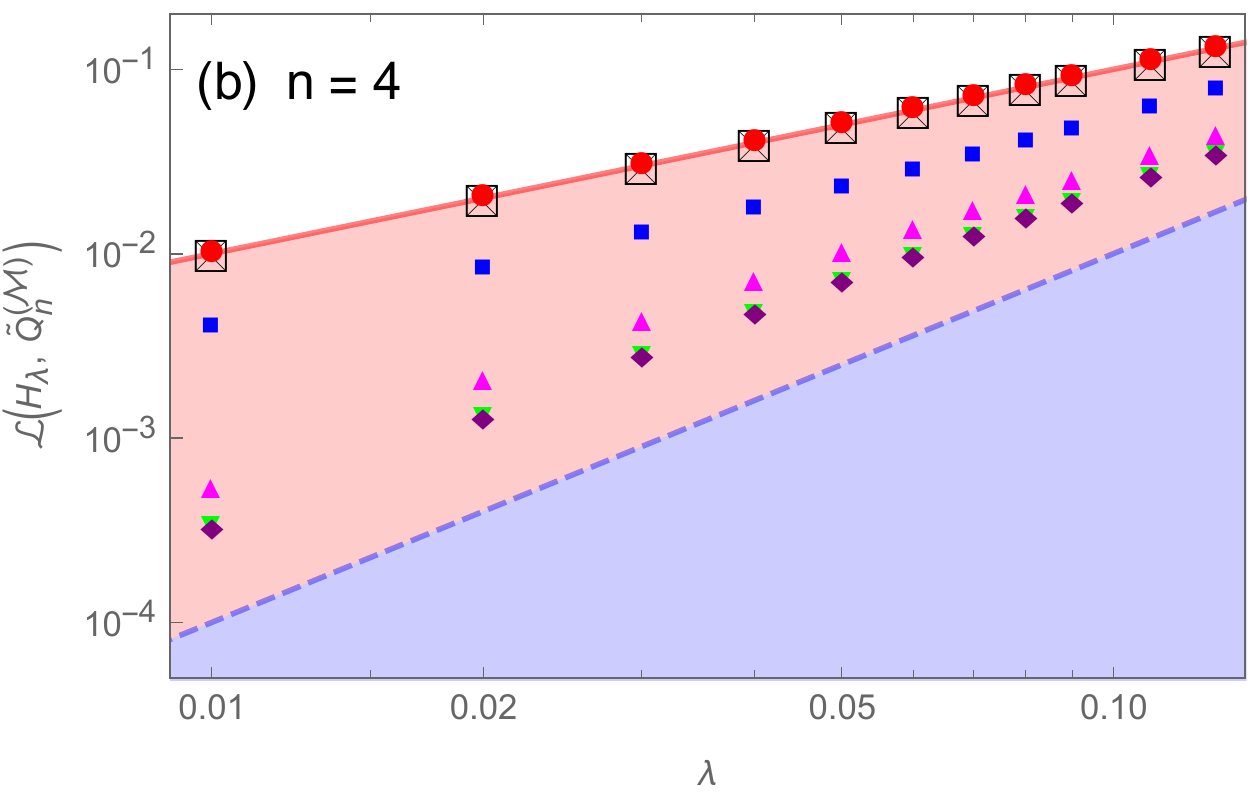}
}\hfill
\subfloat{
  \includegraphics[width=0.49\linewidth]{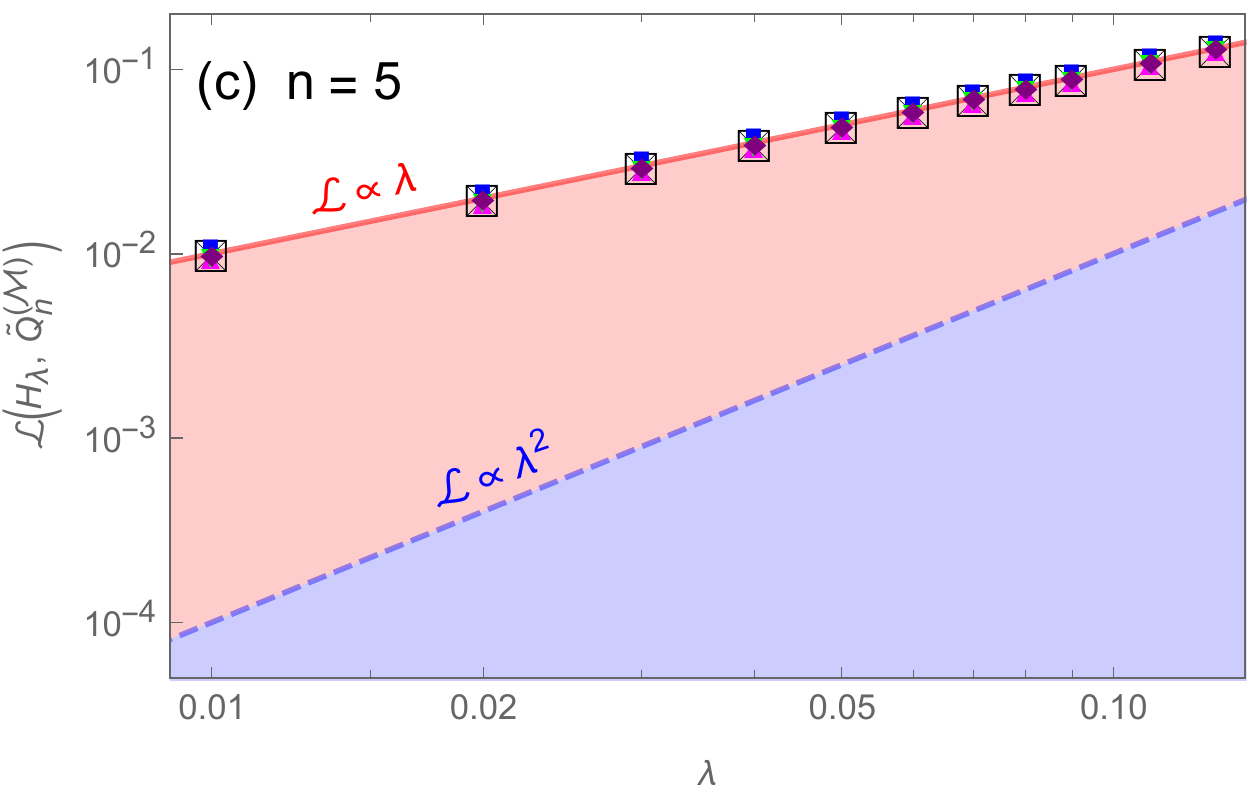}
}\hfill
\subfloat{
  \includegraphics[width=0.49\linewidth]{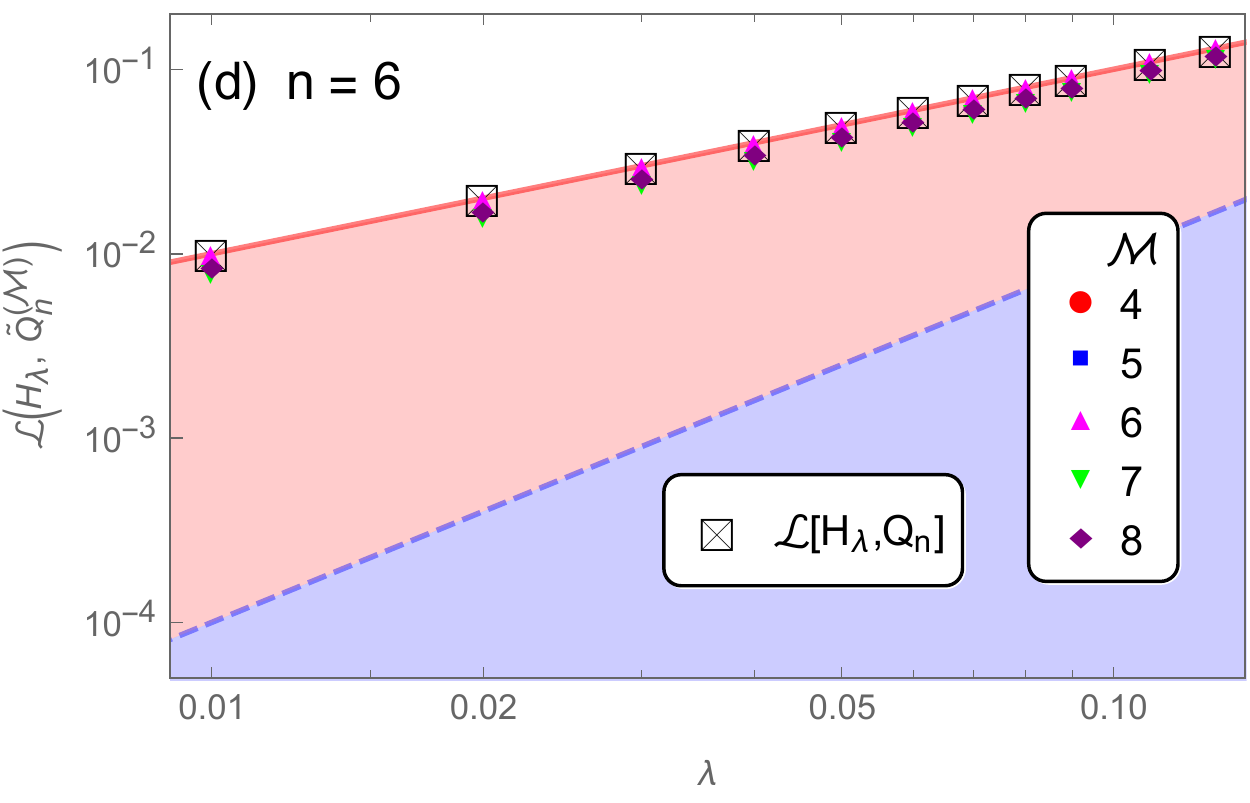}
}
\caption{ Non-commutativity measure ${\cal L}[ H_{\lambda}, \tilde Q_n^{({\cal M})} ]$, defined in Eq.~(\ref{non_commutativity_measure}), versus the perturbation strength $\lambda$ for different quasi-conserved charges $\tilde Q_n^{({\cal M})}$, defined in Eq. (\ref{quasiQ_ansatz}), from $n=3$ to $n=6$ [panels (a) to (d), respectively] and different values of the maximal support ${\cal M}$. In order to focus on the exponent of the power-law scaling, all data points were rescaled such that in Eq.~(\ref{non_commutativity_measure_scaling}) one has $a=1$. Note the log-log scale on all plots. In panels (a)--(d), the symbol $\boxtimes$ shows the values of the non-commutativity measure~${\cal L}[H_{\lambda}, Q_n]$ for the unperturbed charges $Q_n$ (those conserved by $H_0$).  Thus, these data points lie on the line corresponding to the linear scaling (red solid line), as can be easily understood from Eqs.~(\ref{local_charges}), (\ref{H_tot}), and (\ref{non_commutativity_measure}). In panel~(a) we see that the non-commutativity measure for quasi-conserved charge $\tilde Q_3^{({\cal M})}$ behaves as ${\cal L} \propto \lambda^2$ already for the maximal support~${\cal M} = 4$. For all other values of ${\cal M}$ the data points also lie on the line corresponding to quadratic scaling (blue dashed line). Panel~(b) shows that for the next charge $\tilde Q_4^{({\cal M})}$, the non-commutativity measure ${\cal L}$ does not acquire a quadratic behavior even for ${\cal M}=8$, and scales with an exponent $b\approx 1.6$. Finally, in panels (c) and (d) we see that for the charges $\tilde Q_5^{({\cal M})}$ and $\tilde Q_6^{({\cal M})}$, the non-commutativity measure ${\cal L}$ scales linearly with $\lambda$ for all values of ${\cal M}$ that we considered. }
\label{F:scaling}
\end{figure}

\end{widetext}

\section{Results and Discussion}
\label{S:results}

Essentially, Eq.~(\ref{non_commutativity_measure}) gives us the non-commutativity measure $\mathcal{L}[H_{\lambda}, \tilde Q_n^{({\cal M})}]$ as a function of $\lambda$. Since we are dealing with the case of weak perturbations,~$\lambda \ll 1$, the dependence on $\lambda$ comes dominantly from the numerator of $\mathcal{L}$ and coincides with that of the commutator norm~$|| [H_{\lambda}, \tilde Q_n^{({\cal M})} ] ||$. Thus, we have
\be \label{non_commutativity_measure_scaling}
	\mathcal{L}[H_{\lambda}, \tilde Q_n^{({\cal M})} ] \sim a \lambda^b,
\ee 
and according to Eq.~(\ref{comm_norm_scaling}) we need~$b \geq 2$ in order for~$\tilde Q_n^{({\cal M})}$ to be quasi-conserved and survive up to times~$\sim \lambda^{-2}$.

We then calculate the non-commutativity measure $\mathcal{L}[H_{\lambda}, \tilde Q_n^{({\cal M})}]$ for the first four quasi-conserved charges,~$\tilde Q_n^{({\cal M})}$ with $n=3$ to $6$. 
For the maximal support of the $n$-th charge we take the values $n \leq {\cal M} \leq 8$. 
Since we are only interested in the exponent of the power law scaling ${\cal L} \sim a \lambda^b$, we eliminate the numerical coefficient $a$ by fitting the data points to $a \lambda^b$ and rescaling them by $1/a$. 
The results are presented in Fig.~\ref{F:scaling}. One can clearly see that the first non-trivial charge, $\tilde Q_3^{({\cal M})}$, behaves as $\sim \lambda^2$ already at ${\cal M} = 4$. Taking larger values of ${\cal M}$ does not modify this behavior. 
Therefore, $\tilde Q_3^{({\cal M})}$ is indeed a quasi-conserved quantity, which survives during the pre-thermalization phase.  
The situation changes for the next charge, $\tilde Q_4^{({\cal M})}$. We see that as we take larger values of ${\cal M}$, the exponent in $\tilde Q_4^{({\cal M})} \sim \lambda^b$ grows, but remains within the range $1 < b < 2$ even for ${\cal M} = 8$ (in which case $b \approx 1.6$). 
For the next two charges, $\tilde Q_n^{({\cal M})}$ with $n=5$ and~$6$, the picture is again different, as we can see from Figs.~\ref{F:scaling}~(c) and~(d), respectively. There, data points collapse onto the line corresponding to the linear scaling for all values of ${\cal M}$ that we consider. 
This suggests that $\tilde Q_5^{({\cal M})}$ and $\tilde Q_6^{({\cal M})}$ are not even approximately conserved during the pre-thermalization phase. It is then natural to expect that the same holds for all higher-order quantities, $\tilde Q_n^{({\cal M})}$ with $n>6$.

Thus, we come to the conclusion that in a weakly perturbed XXX model (\ref{H_tot}), with the perturbation given by Eq.~(\ref{H_1}), the only quasi-conserved quantity that survives during the pre-thermalization phase up to times~$\sim \lambda^{-2}$ is~$\tilde Q_3^{({\cal M})}$. The next charge $\tilde Q_4^{({\cal M})}$ is marginal, in a sense that it survives up to times $\sim \lambda^{-b}$, with $1 < b < 2$. All higher-order charges die out at much shorter times~$\sim \lambda^{-1}$ and therefore can not play any role in the pre-thermalization phase.
One may argue, that this conclusion is somewhat premature, and by including terms with a sufficiently large maximal support ${\cal M}$, one can make $\tilde Q_4^{({\cal M})} \propto \lambda^2$, and maybe even do so for higher order charges. However, it seems that this possibility is ruled out by the fact that the contribution of $\delta Q_n^{(m)}$ [see Eq.~(\ref{quasiQ_ansatz})] into $\tilde Q_n^{({\cal M})}$ dramatically decreases with $m$. Our detailed analysis shows, that already for $m = 8$, the contribution from $\delta Q_n^{(m)}$ is practically negligible.

We have also performed the same analysis for a perturbation other than that in Eq.~(\ref{H_1}):
\be
	H_1' =\sum_{j} J\, \vec{\sigma}_j\cdot\vec{\sigma}_{j+3}.
\ee
This perturbation is still translationally- and $SU(2)$-invariant, but it is less local, as compared to~$H_1$ in Eq.~(\ref{H_1}). It turns out, that the latter fact plays an important role. Namely, for the Hamiltonian~$H_{\lambda}' = H_0 + H_1'$ we are not able to construct any quasi-conserved quantities. Even for~$\tilde Q_3^{({\cal M})}$ with ${\cal M} = 8$ we obtain that ${\cal L}[H_{\lambda}', \tilde Q_3^{({\cal M})} ]$ scales linearly with~$\lambda$. We conjecture that the situation is the same for any perturbation which is less local than~$H_1$ in Eq.~(\ref{H_1}). Indeed, according to Eqs.~(\ref{H_tot}) and (\ref{quasiQ_ansatz}), in order to eliminate the linear dependence on~$\lambda$ from the commutator $[H_{\lambda},\tilde Q_3^{({\cal M})}]$, one should find a correction $\sum_{m=2}^{{\cal M}} \delta Q_3^{(m)}$ such that the two commutators $[H_0, \sum_m \delta Q_3^{(m)}]$ and $ [ \lambda H_1', Q_3]$ cancel each other. However, it seems that this is problematic when $H_1'$ and $Q_3$ consist of operators with different supports. It might be possible to overcome this problem by considering corrections $\delta Q_3^{(m)}$ with even higher support, but this is beyond the scope of the present paper. 
It is also possible that the stability of integrability (and, consequently, the existence of the prethermalization plateaus) is not a generic property for perturbations of arbitrary nature. If this is the case, then the identification of quasi-conserved charges should be considered only as a sufficient (and not necessary) condition for the prethermalization.

To summarize, in this paper we have shown that an isotropic Heisenberg spin chain, weakly perturbed away from integrability by a next to nearest neighbor interaction of strength $\lambda$, possesses a quasi-conserved charge $\tilde Q_3^{({\cal M})}$ which is approximately conserved up to times of the order $\lambda^{-2}$. The next charge $\tilde Q_4^{(\cal M)}$ survives at shorter times up to $\sim \lambda^{-b}$, with $b\approx 1.6$. The next two charges, $\tilde Q_5^{(\cal M)}$ and $\tilde Q_6^{(\cal M)}$ die out already at times~$\sim\lambda^{-1}$, and similar picture is expected for all higher-order charges. This suggests, that in the prethermal phase only a few quasi-conserved quantities are important. Our findings are supported by the results of Refs.~\cite{EF-rev,EF1,poz1}, which demonstrate that even for the integrable models one can use only a few conserved charges, the most local ones, to reliably approximate the full GGE. This approach is known under the name of truncated GGE (tGGE). Although the studies of the XXZ chain for $\Delta>1$ suggest that the formal convergence of the truncated GGE in the limit $\Delta\rightarrow1$ [which corresponds to the XXX model, given by $H_0$ in Eq.~(\ref{H_0})] requires many integrals \cite{poz1}, the corresponding Lagrange multipliers for higher charges are considerably smaller than for the lowest ones. It is then natural to assume that the truncation does make sense for the perturbed case as well. 
One the other hand, it might be such that one has to consider not only the local charges, but also the quasi-local ones. It is well known that they play an important role in constructing the full GGE for the integrable case  \cite{quasilocal1,quasilocal2, quasilocalgge}. Therefore, including the  quasi-local quasi-conserved integrals of motion can be equally important for the perturbed case. However, this is a subject for future studies.

\section{Acknowledgements}
We would like to thank Anatoli Polkovnikov for igniting this project and for the insightful discussions. We also thank Balasz Pozsgay for drawing our attention to Refs.~\cite{longrange1, longrange2, Pozsgay2020, Marchetto2020} and Anatoli Dymarsky for very useful comments. This work is part of the DeltaITP consortium, a program of the Netherlands Organization for Scientific Research (NWO) that is funded by
the Dutch Ministry of Education, Culture and Science
(OCW).
The results of D.V.K. were supported by the Russian Science
Foundation Grant No. 20-42-05002 (parts of Secs.~\ref{S:model_boost}, \ref{S:quasi_charges}, and~\ref{S:results}).

\appendix
\section{} \label{SU2_inv_basis}

In this appendix we discuss the basis of $SU(2)$-invariant operators used to express~${\cal Q}_j^{(m)}$ in Eq.~(\ref{deltaQ}). The basis is spanned by the $SU(2)$-invariant tensor products of spin-$1/2$ operators on~$k$~sites. Consider an operator that acts non-trivially on sites $j_1, j_2, \ldots, j_k$:
\be
	\sigma^{\alpha_1}_{j_1} \sigma^{\alpha_2}_{j_2} \ldots \sigma^{\alpha_{k}}_{j_{k}}.
\ee
It is well-known that in order to make it $SU(2)$-invariant, one simply has to contract the indices. For example, for the first few values of $k$ one has the following invariants:
\be \label{Delta_k}
\begin{aligned}
	\Delta_2(j_1, j_2) &= -\frac{1}{2} \delta_{\alpha_1 \alpha_2} \sigma_{j_1}^{\alpha_1}\sigma_{j_2}^{\alpha_2}, \\
	\Delta_3(j_1, j_2, j_3) &= \frac{1}{2} \epsilon_{\alpha_1 \alpha_2 \alpha_3}\sigma_{j_1}^{\alpha_1}\sigma_{j_2}^{\alpha_2}\sigma_{j_3}^{\alpha_3},\\
	\Delta_4(j_1,j_2,j_3,j_4) &= -\frac{1}{2} \epsilon_{\alpha_1 \alpha_2 \beta}\epsilon_{\beta \alpha_3 \alpha_4}\sigma_{j_1}^{\alpha_1}\sigma_{j_2}^{\alpha_2}\sigma_{j_3}^{\alpha_3}\sigma_{j_4}^{\alpha_4}, \\
	\Delta_5(j_1,j_2,j_3,j_4,j_5) &= \frac{1}{2} \epsilon_{\alpha_1 \alpha_2 \beta}\epsilon_{\beta \alpha_3 \gamma} \epsilon_{\gamma \alpha_4 \alpha_5} \\
	&\qquad\qquad\times\sigma_{j_1}^{\alpha_1}\sigma_{j_2}^{\alpha_2}\sigma_{j_3}^{\alpha_3}\sigma_{j_4}^{\alpha_4}\sigma_{j_5}^{\alpha_5},
\end{aligned}
\ee
where $\delta_{\alpha_i \alpha_j}$ is the Kronecker symbol,  $\epsilon_{\alpha_i \alpha_j\alpha_k}$ is the Levi-Civita tensor, $\alpha_i \in \{1,2,3\}$, and the summation over repeating indices is implied. Thus, $\Delta_m$ is a sum of $3\times 2^{m-2}$ terms. The factors of $\pm1/2$ are included for later convenience.
Note, that one may construct more invariants by contracting other indices. However, our detailed analysis shows that including these extra terms does not improve the exponent in Eq.~(\ref{non_commutativity_measure_scaling}) and only slightly modifies the numerical prefactor. This might be related to the fact the basis of all SU(2)-invariant operators is overcomplete~\cite{Lychcovskiy2018,su2}.

One can show that for the operators defined in Eq.~(\ref{Delta_k}) the following recursive relation holds:
\be
	\Delta_n(j_1, \ldots, j_n) = i \left[ \Delta_2(j_1, j_2), \Delta_{n-1}(j_2,  \ldots, j_{n}) \right],
\ee
which is valid if all $j_k$ are distinct. This relation allows one to generate the whole basis of $SU(2)$-invariant tensor products of spin-$1/2$ operators that is used to generate the  corrections $\delta Q_n^{(m)}$ in Eq.~(\ref{deltaQ}). 
As an illustration, let us give the explicit form for the first few corrections:
\be
\begin{aligned}
	\delta Q_n^{(2)} &= \sum_j c_n^{(2)}\left( \{ 0,1\} \right) \Delta_2 (j, j+1), \\
	\delta Q_n^{(3)} &= \sum_j \Bigl[ c_n^{(3)}\left( \{ 0,2\} \right) \Delta_2 (j, j+2) \\
				&+ c_n^{(3)}\left( \{ 0,1,2\} \right) \Delta_3 (j, j+1, j+2) \Bigr], \\
	\delta Q_n^{(4)} &= \sum_j\Bigl[ c_n^{(4)}\left( \{ 0,3\} \right) \Delta_2 (j, j+3) \\
				&+  c_n^{(4)}\left( \{ 0,1,3\} \right) \Delta_3 (j, j+1, j+3) \\
				&+  c_n^{(4)}\left( \{ 0,2,3\} \right) \Delta_3 (j, j+2, j+3) \\
				&+  c_n^{(4)}\left( \{ 0,1,2,3\} \right) \Delta_4 (j, j+1, j+2, j+3) \Bigr]. 		
\end{aligned}
\ee
For $n = 3$ these operators, together with Eq.~(\ref{Q3}), yield the quasi-conserved charge $\tilde Q_3^{(4)} $ [see Eq.~(\ref{quasiQ_ansatz}) and~Fig.~\ref{F:scaling}~(a)] and we need to put $c_{3}^{(3)}(\{ 0,1,2\}) = 0$, as discussed after~Eq.~(\ref{sites}).

\end{document}